\documentclass[pdflatex,sn-mathphys-ay]{sn-jnl}

\usepackage{graphicx}%
\usepackage{multirow}%
\usepackage{amsmath,amssymb,amsfonts}%
\usepackage{amsthm}%
\usepackage{mathrsfs}%
\usepackage[title]{appendix}%
\usepackage{xcolor}%
\usepackage{textcomp}%
\usepackage{manyfoot}%
\usepackage{booktabs}%
\usepackage{algorithm}%
\usepackage{algorithmicx}%
\usepackage{algpseudocode}%
\usepackage{listings}%
\usepackage{amssymb}
\usepackage{bm}
\usepackage{subcaption}

\theoremstyle{thmstyleone}%
\theoremstyle{thmstyletwo}%

\theoremstyle{thmstylethree}%

\begin{document}

\title[Spillovers and Co-movements in Multivariate Volatility: A Vector Multiplicative Error Model]{Spillovers and Co-movements in Multivariate Volatility: A Vector Multiplicative Error Model}


\author*[1]{\fnm{Edoardo} \sur{Otranto}}\email{edoardo.otranto@uniroma1.it}\equalcont{These authors contributed equally to this work.}

\author[2]{\fnm{Luca} \sur{Scaffidi Domianello}}\email{luca.scaffidi@unict.it}
\equalcont{These authors contributed equally to this work.}

\affil[1]{\orgdiv{Department of Social and Economic Sciences}, \orgname{University Sapienza-Rome, and CRENoS}, \orgaddress{\street{Piazzale Aldo Moro, 5}, \city{Rome}, \postcode{00185}, \country{Italy}}}

\affil[2]{\orgdiv{Department of Economics and Business}, \orgname{University of Catania}, \orgaddress{\street{Corso Italia, 55}, \city{Catania}, \postcode{95122}, \country{Italy}}}


\abstract{Recent developments in financial time series focus on modeling volatility across multiple assets or indices in a multivariate framework, accounting for potential interactions such as spillover effects. Furthermore, the increasing integration of global financial markets provides a similar dynamics (referred to as co-movement). In this context, we introduce a novel model for volatility vectors within the Multiplicative Error Model (MEM) class. This framework accommodates both spillover and co-movement effects through a distinct latent component. By adopting a specific parameterization, the model remains computationally feasible even for high-dimensional volatility vectors.  To reduce the number of unknown coefficients, we propose a simple model-based clustering procedure. We illustrate the effectiveness of the proposed approach through an empirical application to 29 assets of the Dow Jones Industrial Average index, providing insight into volatility spillovers and shared market dynamics. Comparative analysis against alternative vector MEMs, including a fully parameterized version of the proposed model, demonstrates its superior or at least comparable performance across multiple evaluation criteria.}




\maketitle

\section{Introduction}\label{sec:intro}
The growing globalization of financial markets and their increasing interdependence have favored the development of multivariate models for vectors of volatility indices or financial assets. Early contributions in this area extended the classical GARCH framework \citep{Bollerslev86} to multivariate contexts for the modeling of covariance matrices, such as the BEKK model by \citet{Engle_Kroner95}, and dynamic correlation structures, as exemplified by the DCC model of \citet{EngleDCC02}. A central challenge in modeling such covariance or correlation matrices lies in specifying parameterizations that not only ensure positive definiteness but also maintain parsimony, thereby mitigating the so-called {\it curse of dimensionality} \citep{BLRsurvey06}.  Commonly adopted solutions rely on scalar or diagonal structures, which rule out dependence of each conditional variance on lagged variances of other variables. Alternatively, the two-step estimation procedure of the DCC model treats conditional variances via independent univariate GARCH models. As a result, traditional multivariate GARCH-type models generally preclude the incorporation of spillover effects in their formulations.

A more recent and flexible framework is the Multiplicative Error Model (MEM) proposed by \citet{Engle_NewF_2002}, which models non-negative time series through the product of two factors, representing the conditional mean and a strictly positive innovation term, respectively. The success of MEM and its extensions is partly due to the increasing availability of more accurate volatility measures, including the high-low range-based volatility \citep[HLR;][]{Parkinson1980} and realized volatility (RV) derived from ultra-high-frequency data \citep{AndersenBoll98, ABDL01_Econometrica}, with refinements that address microstructure noise \citep{BN-etal2008}. Both indicators exploit intraday price variations: the HLR is derived from the daily maximum-minimum spread and is straightforward to compute, whereas RV requires multiple intraday observations. Although RV is widely regarded as a more accurate volatility measure, its application is often limited by data availability. HLR, on the contrary, is less information-intensive but broadly available across indices and individual assets. Both metrics are relatively robust to microstructure noise and generally superior to traditional GARCH-based estimates that rely on squared or absolute returns \citep{Chou:Chou:liu:2015}.

The MEM framework is particularly well suited to modeling such non-negative series. Its multivariate extension, the vector MEM (vMEM), was introduced by \citet{EngleGallo_2006}, wherein the dependent variable is a vector of positive-valued time series. Their empirical application considers a 3-variate vector comprising absolute returns, daily ranges, and realized volatilities. They impose the strong assumption of mutual independence among innovation terms, enabling an equation-by-equation estimation strategy. A key advancement in this area was offered by \citet{Cipollini:Engle:Gallo:2013}, who developed a semiparametric model, avoiding specific distributional assumptions, and proposed a GMM estimator capable of capturing interdependence across the variables. Subsequently, \citet{Cipollini:Engle:Gallo:2017} employed a copula-based framework to model dependence between Gamma and log-normal innovation marginal distributions, showing its empirical equivalence to the semiparametric approach. A notable refinement is proposed by \citet{Tailor:Xu:2017}, who specify log-normal innovations and model the logarithm of the conditional mean, avoiding the need to impose positivity constraints on parameters. Through simulation experiments, they demonstrate the consistency of this method relative to the semiparametric estimator of \citet{Cipollini:Engle:Gallo:2013}. These approaches have primarily been applied to 3-variate systems to avoid high-dimensional complications; indeed, even a fully parameterized 3-dimensional vMEM involves 24 parameters, increasing to 42 in four dimensions.

In this paper, we propose a novel diagonal log-vMEM, building upon the work of \citet{Tailor:Xu:2017}, augmented by an additive common component that captures long-term market co-movements and volatility spillover effects. This common factor exhibits autoregressive dynamics and is derived from the first principal component of the panel of volatilities, thereby including both common temporal dynamics and lagged spillover effects. A related approach is found in \citet{Atak_Kapetanios_2013}, who enhance the Heterogeneous AutoRegressive (HAR) model of \citet{Corsi_09} by including principal component effects. In our specification, the co-movement factor is isolated from the standard GARCH-type conditional mean equation, allowing us to disentangle long-run common volatility components from asset-specific idiosyncrasies. This decomposition of the conditional mean into two additive terms can be viewed as a multivariate extension of the Spillover Asymmetric MEM (SAMEM) introduced by \citet{Otranto:2015}, which itself belongs to the broader class of Composite MEMs \citep{mem_hbv_2012}. To address the challenges posed by high-dimensional volatility vectors, we propose a reparameterization of the model coefficients based on a model-based clustering algorithm. We apply this new modeling strategy to a panel of 29 assets from the Dow Jones Industrial Average (DJIA) and compare alternative parameterizations of the proposed model. The empirical analysis reveals enhanced performance and interpretability of the reduced-parameter specification, despite the substantial dimensionality reduction.

The remainder of the paper is structured as follows. Section 2 introduces the proposed model. Subsection \ref{sec:un_model} details the fully parameterized version, subsection \ref{sec_constr} discusses its statistical properties and the resulting parameter constraints, and subsection \ref{sec:con_model} presents the proposed parameterization suited to high-dimensional data. Section \ref{sec:apps} reports the empirical application to 29 DJIA assets, beginning with the dimensionality reduction strategy, followed by model estimation and interpretation of results, including comparisons with standard vMEMs. Subsection \ref{sec:fac} provides a graphical analysis of the effects of the common component. Final remarks are presented in the concluding section.

\section{The model proposed}\label{sec:model}
Let $\bm{y}_t$ denote a vector of $n$ volatility indices (assets) observed at time $t$. We specify a vMEM, following the framework of \citet{Cipollini:Engle:Gallo:2017}, augmented to incorporate both spillover effects through index-specific parameters and an unobservable common component capturing co-movements across all series. To facilitate estimation, we apply a logarithmic transformation to the variables and adopt the log-vMEM formulation introduced by \citet{Tailor:Xu:2017}. This log-specification simplifies the estimation procedure by enabling the use of a log-normal-based likelihood, while also allowing for a straightforward integration of the common component into the conditional mean structure.

\subsection{The fully parameterized model}\label{sec:un_model}
Let ${\bm x}_t$ denote the logarithm of ${\bm y}_t$. The proposed model is specified as follows:
\begin{align}
	\begin{split}
		{\bm y}_t&={\bm \mu}_t \odot {\bm \varepsilon}_t   \qquad {\bm \varepsilon}_t \sim {\ln N}({\bm m},{\bm V})  \\
		\ln {\bm \mu}_t&={\bm \varsigma}_t + \bm{\vartheta}\xi_t  \\
		\xi_t&=  \delta	p_{t-1} + \phi\xi_{t-1} \\
		{\bm\nu_t}&= {\bm x_t}-\bm{\vartheta}\xi_t \\
		{\bm \varsigma}_t&= ({\bm I_n} - {\bm{A}}- {\bm{B}}) \bar{{\bm x}} + ({\bm I_n} - {\bm{B}})diag(\bm{V})/2+ {\bm A} {\bm \nu}_{t-1}+{\bm B}  {\bm \varsigma}_{t-1} .
	\end{split}\label{vmem_sc}	
\end{align} 

Here, $\odot$ denotes the element-wise (Hadamard) product and ${\bm I}_n$ is the $n \times n$ identity matrix. Following  \citet{Tailor:Xu:2017}, the innovation vector ${\bm \varepsilon}_t$ is log-normally distributed with parameters ${\bm m} = (m_1,\dots,m_n)'$ and $\bm{V}=\left\{v_{ij}\right\}$ ($i,j=1,\dots,n$),  a symmetric and positive definite $n \times n$ covariance matrix. To ensure unit mean for each component of ${\bm \varepsilon}_t$, we impose the constraint $m_i=-v_{ii}/2$ ($i=1,\dots,n$). As a result, ${\bm y}_t\sim {\ln N}(\ln {\bm \mu}_t +{\bm m},{\bm V}) $. The vector $\bar{\bm x}$ denotes the sample mean of the log-volatility vector ${\bm x}_t$.

In (\ref{vmem_sc}) $\ln {\bm \mu}_t$ is decomposed into the sum of two unobservable components:  the idiosyncratic log-volatility vector ${\bm \varsigma}_t$ and a common dynamic component $\xi_t$, which has zero mean and enters each element of $\ln {\bm \mu}_t$ with a factor loading $\vartheta_i$, collected in the vector $\bm{\vartheta}$. To avoid identification problems stemming from the scaling indeterminacy between $\bm{\vartheta}$ and $\delta$, we normalize $\bm{\vartheta}$ such that $\bm{\vartheta}' \bm{\iota}_n = n$, with ${\bm \iota}_n$ an $n$-dimensional column vector of ones, as suggested by \citet{Cipollini:Gallo:2019}.

The unobservable common time-varying component $\xi_t$ evolves as an AR(1) process, driven by its own lag and by $p_{t-1}$, the first principal component of the $n$ (demeaned) log-volatilities in $\bm{x}_t$. Although it is possible to include more principal components to capture a larger percentage of variance explained, the high degree of co-movement among volatility series often makes the first component sufficient, as it typically explains more than 50\% of the total variance.

The vector ${\bm \varsigma}_t$ captures the idiosyncratic dynamics of each series net of the common component $\xi_t$ and follows a multivariate GARCH-type structure. We assume ${\bm A}$ and ${\bm B}$ to be diagonal matrices, implying that each component of ${\bm \varsigma}_t$ depends only on its own lagged value and the corresponding ${\bm \nu}_{t-1}$.
This structure exploits the concept of expectation targeting for multivariate volatility models \citep{Laurent:Rombouts:Violante:2012, Cipollini:Engle:Gallo:2017}, whereby the intercept term is parameterized as $({\bm I_n} - {\bm{A}}- {\bm{B}}) \bar{{\bm x}} + ({\bm I_n} - {\bm{B}})diag(\bm{V})/2$. 
The derivation of the expectation targeting specification is illustrated in Appendix~\ref{targeting}; it implies that the unconditional expectation of ${\bm \varsigma}_t$ is $\bar{{\bm x}}+diag(\bm{V})/2$. Furthermore, the expectation of $\xi_t$, assumed stationary, is equal to 0, since, by construction, each component has zero mean, i.e. $E(p_t)=0$ \citep[see, e.g.,][]{Mardia:1979}. These results imply that the unconditional expectation of $\ln \bm{\mu}_t$ is the same as that of ${\bm \varsigma}_t$.

Thus, the model captures three key components of each log-volatility: individual dynamics, spillover effects, and a common factor reflecting co-movements.  To better illustrate these features,  consider the $i-$th element of $\ln{\bm \mu}_t$.  
Since $p_t=\bm{c}'(\bm{x}_t-\bar{\bm{x}})$ is a linear combination  of the $n$  log-volatilities in $\bm{x}_t=(x_1,x_2,\dots x_n)'$, with weights $\bm{c}=(c_1,c_2,\dots,c_n)'$, and  developing the $i-$th equation of $\bm{\varsigma}_t$, we can rewrite, after simple algebra, the $i$-th conditional mean as :
\begin{equation}
	\ln \mu_{i,t}=\bar{x}_i+(\alpha_i +\vartheta_i  \delta c_i)(x_{i,t-1}-\bar{x}_i)+\beta_i (\ln \mu_{i,t-1}-\bar{x}_i)+\sum_{j \ne i} \vartheta_i \delta c_j (x_{j,t-1}-\bar{x}_j)+(\phi-\alpha_i-\beta_i)  \vartheta_i\xi_{t-1} \label{unimu}
\end{equation} 
Here, $\alpha_i$ and $\beta_i$ are the diagonal elements of matrices ${\bm A}$ and ${\bm B}$, respectively. The first part of equation (\ref{unimu}) corresponds to the standard MEM structure, while the term $ \theta_i \delta c_j$ captures the spillover from variable $j$ to $i$ (different for each pair $(i,j)$). The last term, involving the lagged common factor, reflects dynamic co-movement effects with coefficient $( \phi-\alpha_i-\beta_i )\vartheta_i$. Prior literature has shown that co-movement is closely related to volatility spillovers  \citep[see, for example,][and its references]{Chai2020}. Because $\xi_t$ depends on $p_{t-1}$ and past values of itself, the model can separately identify {\it pure} spillover effects and common component effects,  which depends on the same spillover and its lagged value (third equation of (\ref{vmem_sc})). 

We refer to the proposed specification in (\ref{vmem_sc}) as the vMEM with Spillover Effects and Co-movement (vMEM-SeC).

\subsection{Statistical Properties}\label{sec_constr}
A key advantage of modeling volatility in logarithmic form is that it removes the need to impose positivity constraints on the model parameters.  Stationarity and invertibility conditions are determined solely by the third and fifth equations in system (\ref{vmem_sc}).
For the common component $\xi_t$, by adding and subtracting $p_t$ and $\phi p_{t-1}$ in the third equation of (\ref{vmem_sc}), it is possible to rewrite the process as an ARMA(1,1) specification for $p_t$:
\begin{equation}
	p_t=(\delta+\phi)p_{t-1} - \phi (p_{t-1}-\xi_{t-1}) +  (p_{t}-\xi_{t})\label{eq1}
\end{equation}
where $(p_{t}-\xi_{t})$ represents the innovation term. Consequently, the autoregressive (AR) coefficient is $(\delta + \phi)$, and the moving average (MA) coefficient is $-\phi$. Stationarity and invertibility of the ARMA process require the following constraints:
\begin{align}
	\begin{split}
		\lvert \delta+ \phi\rvert &< 1 \\	
		\lvert \phi \lvert&<1
	\end{split}\label{con1}
\end{align}
Regarding the idiosyncratic component ${\bm \varsigma}_t$, stationarity conditions can be derived from Appendix A of \citet{Tailor:Xu:2017}. The process is stationary if the roots of the characteristic equations:
\begin{align*}
	\begin{split}
		|\bm{I}_n-( {\bf A}z+\bm{B}z)|&=\prod_{i=1}^n \left[1-(\alpha_i+\beta_i) z\right]=0\\
		|\bm{I}_n-( \breve{\bf A}z+\bm{B}z)|&=\prod_{i=1}^n \left[1-(\alpha_i+ \delta c_i+\beta_i) z\right]=0
	\end{split}
\end{align*}
lie outside the unit circle. The first expression derives from the fifth equation of system (\ref{vmem_sc}), while the second originates from equation (\ref{unimu}), where $\breve{\bf A}$ is a diagonal matrix with elements $(\alpha_i + \delta c_i)$. Accordingly, the stationarity conditions for each $i = 1, \dots, n$ are:
\begin{align}
	\begin{split}
		(\alpha_i+\beta_i)&<1\\
		\delta c_i&<1-(\alpha_i+\beta_i)\\
	\end{split} \label{con2}
\end{align}
Invertibility requires that the roots of the following polynomial lie outside the unit circle:
\begin{equation*}
	|\bm{I}_n-\bm{B}z|=\prod_{i=1}^n (1-\beta_i z)=0
\end{equation*}
This implies the following condition for each $i$:
\begin{equation}
	\vert \beta_i \vert <1  \label{con3}
\end{equation}
Estimation of the model parameters is carried out via Maximum Likelihood Estimation (MLE). Letting $\bm{\theta}$ denote the vector of parameters to be estimated, and assuming log-normality of the innovations, the log-likelihood function under constraints (\ref{con1})-(\ref{con3}) is given by:
\begin{equation}
	l(\bm{\theta})=-\frac{Tn}{2}\ln 2\pi-\frac{T}{2}\ln |\bm{V}|-\sum_{t=1}^{T}\left[\sum_{i=1}^{n}x_{i,t}-\frac{1}{2}\left(\bm{x}_t-\ln \bm{\mu}_t-\bm{m}\right)' \bm{V}^{-1}\left(\bm{x}_t-\ln \bm{\mu}_t-\bm{m}\right)\right] \label{loglik}
\end{equation}	
Robust standard errors can be obtained using a sandwich estimator of the covariance matrix, as proposed by \citet{White82}. Asymptotic properties of the MLE in the context of the log-vMEM framework are discussed in detail by \citet{Tailor:Xu:2017}.

\subsection{Parameterization for large datasets}\label{sec:con_model}
The total number of unknown parameters in (\ref{vmem_sc}), including the elements of the covariance matrix $\bm{V}$, is given by $n(n+7)/2+1$. As the dimension $n$ increases, this parameter space grows rapidly, leading to the well-known issue of the curse of dimensionality. Consequently, it becomes essential to adopt a suitable parameterization strategy that ensures both parsimony and feasibility of estimation. One approach to reduce the number of parameters involves identifying groups of series that exhibit similar dynamic behavior, such that they can share common parameters $\alpha_i$ and $\beta_i$ (in the final equation of system \eqref{vmem_sc}) and common loading coefficients $\vartheta_i$ (in the second line). In particular, the grouping of (${\alpha_i,\beta_i}$) and ${\vartheta_i}$ does not need to coincide and is identified in separate steps. This reparameterization can be operationalized through a straightforward algorithm, beginning with a preliminary estimate of the unobservable common component $\xi_t$.

The algorithm begins with a simplified version of model (\ref{vmem_sc}), where matrices $\bm{A}$ and $\bm{B}$ are restricted to be scalar. Estimating this model yields a preliminary estimate of the common component, denoted $\xi_t^*$. 
This preliminary estimate is then used as a regressor in a second-stage estimation based on a univariate MEM-SeC model, essentially, the univariate analogue of model (\ref{vmem_sc}):
\begin{align}
	\begin{split}
		y_{i,t}&=\mu_{i,t} \varepsilon_{i,t}     \ \qquad \varepsilon_{i,t} \sim {\ln N}(m_i,v_{ii}) \ \quad \forall i=1,\dots,N \\
		\ln \mu_{i,t}&=\varsigma_{i,t} + \vartheta_i\xi_t^*  \\
		\nu_{i,t}&= x_{i,t}-\vartheta_i\xi_t^* \\
		\varsigma_{i,t}&= (1 - \alpha_i- \beta_i) \bar{x_i} + (1 - \beta_i)v_{ii}/2+  \alpha_i \nu_{i,t-1}+\beta_i  \varsigma_{i,t-1} \\
	\end{split}
	\label{unimem}
\end{align}  
where $m_i = -v_{ii}/2$ ensures that ${E}(\varepsilon_{i,t}) = 1$.

From each univariate model, we recover parameter estimates ($\alpha_i^*,\beta_i^*$). These are then used to implement a model-based clustering procedure for the idiosyncratic components of model (\ref{vmem_sc}), grouping series that follow similar GARCH-type dynamics.  A widely used metric for such clustering is the autoregressive (AR) distance proposed by \citet{Piccolo_1990}, which is based on the Euclidean distance between the parameters of the infinite-order AR representations of two invertible processes. In the MEM case, this reduces to a closed-form {\it ARMA distance} between series $i$ and $j$:\footnote{See Appendix \ref{app:dist} for derivation in the general ARMA(1,1) case.}
\begin{equation}
	d_{ARMA}=\left[\frac{\alpha_{i}^{*2}}{1-\beta_i^{*2}}+\frac{\alpha_{j}^{*2}}{1-\beta_j^{*2}}-2\frac{\alpha_{i}^*\alpha_{j}^*}{1-\beta_i^* \beta_j^*} \right]^{1/2} \label{armadist}
\end{equation}
This clustering yields $k_1$ groups, each defined by a distinct pair $(\alpha, \beta)$, denoted $(\alpha_1, \dots, \alpha_{k_1}, \beta_1, \dots, \beta_{k_1})$.
In parallel, we estimate the loading parameters $\vartheta_i^*$ from the univariate MEM-SeC models.\footnote{Since $\xi_t^*$ is treated as a known regressor in this step, no identification constraint is needed on $\vartheta$.} Clustering of ${\vartheta_i^*}$ into $k_2$ homogeneous groups can then be performed using the simple Euclidean distance:
\begin{equation}
	d_{\vartheta}=\left|\vartheta_i^*-\vartheta_j^*\right| \label{dtheta}
\end{equation}

The overall algorithm proceeds as follows:
\begin{enumerate}
	\item Estimate the scalar version of equation \eqref{vmem_sc};
	\item Derive $\xi_t^*$ from the estimated model in step 1.;
	\item Estimate a univariate MEM-SeC model for each series $x_{i,t}$ as in equation (\ref{unimem}), yielding ($\alpha_i^*,\beta_i^*,\vartheta_i^*$);
	\item Cluster the $(\alpha_i^*, \beta_i^*)$ pairs into $k_1$ groups using the ARMA distance in equation (\ref{armadist});
	\item Cluster the $\vartheta_i^*$ values into $k_2$ groups using the distance in (\ref{dtheta}).	
	
\end{enumerate}

Given the identification constraint $\bm{\vartheta}' \bm{\iota}_n = n$ in the full model, only $k_2-1$ distinct $\vartheta$ coefficients must be estimated directly. The vector $\bm{\vartheta}$ thus contains only $k_2$ unique elements.  

Another computational challenge arises in estimating the covariance matrix $\bm{V}$, which contains $n(n+1)/2$ parameters. In addition to dimensionality concerns, ensuring positive definiteness is critical.  
To address both, we adopt the iterative estimation strategy proposed by \citet{Cipollini:Engle:Gallo:2017}, using the sample covariance of the residuals $(\bm{x}_t - \ln \hat{\bm{\mu}}_t)$ as a consistent estimator of $\bm{V}$. The procedure involves the following steps:\footnote{The hat indicates the estimated value.}

\begin{itemize}
	\item[a)] Initialize the covariance matrix $\bm{\hat{V}}$ (for example, putting it equal to the sample covariance matrix) and set $\bm{m} = diag(\bm{\hat{V}})/2$;
	\item[b)] estimate $\bm{{\theta}}=(diag(\bm A),diag(\bm B), \bm{\vartheta}', \phi,\delta)'$ by maximizing the log-likelihood function in equation \eqref{loglik} at the current estimate of $\bm{V}$.
	\item[c)] Compute the log-residuals $\ln\hat{\bm{\epsilon}}_t= (\bm{x}_t-\ln \hat{\bm{\mu}}_t), (t = 1, . . . , T)$, at the current parameter estimates contained in the vector $\hat{\bm{{\theta}}}$;
	\item[d)] Estimate the sample covariance matrix of $\ln\hat{\bm{\epsilon}}_t$ (call it $\bm{\hat{V}}$) and set $\bm{m} = diag(\bm{\hat{V}})/2$;
	\item[e)] Repeat steps b) - d) until convergence, defined as the absolute change of two consecutive maximized log-likelihood falling below a pre-specified threshold.\footnote{In the application of Section \ref{sec:apps} the threshold is set to 1e-4.}
\end{itemize}

The combined use of clustering for the components of $\bm{A}$, $\bm{B}$, and $\bm{\vartheta}$ and the iterative concentration of the log-likelihood function with respect to $\bm{V}$ results in a substantial reduction in the number of parameters. Specifically, the total number of parameters in the reparameterized model is $p = 2(k_1 + 1) + (k_2 - 1)$. For example, when $n = 10$ and the clustering yields $k_1 = k_2 = 4$, the fully parameterized model involves 86 parameters, while the proposed approach reduces this to just 13.

\section{An Empirical Illustration: Components of the Dow Jones Industrial Average}\label{sec:apps}

The dataset comprises 29 of the 30 constituent stocks of DJIA.\footnote{Dow Inc. (DOW) excluded due to its shorter time series, being listed only since March 20, 2019.} The selected assets include: 
Apple Inc. (AAPL),	Amgen Inc. (AMGN),	Amazon.com Inc. (AMZN),	 American Express Company (AXP),	The Boeing Company (BA), Caterpillar Inc. (CAT), Salesforce Inc. (CRM),	Cisco Systems Inc. (CSCO),	Chevron Corporation (CVX),	 The Walt Disney Company (DIS),	 The Goldman Sachs Group Inc. (GS),	 The Home Depot  Inc. (HD),		Honeywell International Inc. (HON),	 International Business Machines (IBM),  Intel Corporation (INTC),	Johnson \& Johnson (JNJ),  JPMorgan Chase \& Co. (JPM),	 The Coca-Cola Company (KO),	McDonald's Corporation (MCD),  3M Company (MMM), Merck \& Co. Inc. (MRK),	 Microsoft Corporation (MSFT),	 NIKE Inc. (NKE),	 The Procter \& Gamble Company (PG), The Travelers Companies Inc. (TRV),	 United Health Group Inc. (UNH), Visa Inc. (V),	 Verizon Communications Inc. (VZ),	 Walmart Inc. (WMT).

Daily data for each asset were retrieved from Yahoo Finance, covering the period from March 19, 2008, to April 22, 2024. For each stock, we construct a proxy for daily volatility based on  HLR as proposed by \citet{Parkinson1980}. Specifically, the volatility proxy is defined as the squared difference between the natural logarithms of the daily high and low prices, rescaled by the factor $4\ln(2)$ and multiplied by 100 for normalization.

The full sample consists of 4,051 daily observations per asset and is initially used for in-sample model estimation and comparison. To assess predictive performance, all models are subsequently re-estimated using data up to December 2022, and the remaining 327 observations (from January 2023 to April 2024) are reserved for out-of-sample evaluation.

\subsection{Estimation and Comparison}\label{sec:est}
We consider six competing models belong to the vMEM-SeC family with three alternative parameterizations: 

\begin{enumerate}
	\item {\it scalar parameterization} (s-vMEM-SeC): in (\ref{vmem_sc}) $\bm{A}$ and  $\bm{B}$  are scalar.
	\item {\it diagonal parameterization} (d-vMEM-SeC): in (\ref{vmem_sc}) $\bm{A}$, $\bm{B}$ are diagonal matrices, 
	each containing $n$ unknown coefficients.
	\item {\it clustered parameterization} (c-vMEM-SeC) : applying the algorithm described in subsection \ref{sec:con_model}, $\bm{A}$ and $\bm{B}$ are diagonal matrices with $k_1$ distinct unknown coefficients, and $\bm{\vartheta}$ is a vector with $k_2$ distinct elements.
\end{enumerate}

We also consider the classical vMEM:
\begin{equation}
	\begin{array}{l}
		{\bm y}_t={\bm \mu}_t \odot {\bm \varepsilon}_t   \qquad {\bm \varepsilon}_t \sim {\ln N}({\bm m},{\bm V})   \\
		\ln {\bm \mu}_t= ({\bm I_n} - {\bm{A}}- {\bm{B}}) \bar{{\bm x}}+(\bm{I_n}-\bm{B})diag(\bm{V}/2)+{\bm A} {\bm x}_{t-1}+{\bm B}  \ln {\bm \mu}_{t-1} 
	\end{array}\label{vmem}
\end{equation} 

We apply the three aforementioned parameterizations also to the vMEM model, yielding s-vMEM, d-vMEM, and c-vMEM.

All models are estimated using the iterative procedure described in subsection \ref{sec:con_model}, avoiding additional unknown coefficients in the covariance matrix $\bm{V}$. The most complex specification, d-vMEM-SeC, involves 32 unknown coefficients (compared to 20 for d-vMEM).

\begin{table}[t!]
	\centering
	\caption{Parameter clusters identified for the c-vMEM-SeC model. Parameters in the diagonals of $\bm{A}$ and $\bm{B}$ are grouped into four clusters (labeled 1 to 4), and $\bm{\vartheta}$ into four clusters (also labeled 1 to 4). Corresponding clusters for c-vMEM (labeled 1 to 3) are indicated in parentheses.\label{tab:clust}}
	{\begin{tabular}{c|cccccccccc} \hline
			asset	&AAPL& AMGN& AMZN&  AXP&   BA&  CAT&  CRM& CSCO&  CVX&  DIS\\
			$(\alpha, \beta)$& 1 (1) &   2 (2) &   1 (3)  &   3 (3)  &  3  (2) & 2   (2)  &  2  (3)  &  3  (2)  &  2  (1)  &   2 (3)   \\
			$\vartheta$& 1 	&    1&    1&    2&    3&    1&    3 &   3&    2&    1  \\
			asset      &GS&   HD&  HON&  IBM& INTC&  JNJ&  JPM&   KO&  MCD&  MMM\\
			$(\alpha, \beta)$& 2  (3)&  2   (3)&   4 (3)  &    2  (3)&   3 (3)  & 2   (2) & 3 (3)&   3  (2)&   3  (3)&   3  (2)\\
			$\vartheta$& 3 
			&    3&    4&    3&    3&    3&    2 &   2&    2&    3  \\
			asset      &MRK& MSFT&  NKE&   PG&  TRV&  UNH&    V&   VZ&  WMT& \\
			$(\alpha, \beta)$&3 (2)&    3 (3)&   3  (2)&   3  (2)& 3    (3)&  3   (2)&     2 (3)&  3   (3) &   2 (2) &\\
			$\vartheta$& 3 
			&    2&    2&    2&    2&    1&    2 &   2&    1&     \\
			\hline
	\end{tabular}}
\end{table}
The estimation of all three vMEM-SeC models requires, as a preliminary step, the first principal component $p_t$ of the 29 time series, which explains 57\% of the total variance, indicating a strong co-movement among the series. For c-vMEM-SeC, the clustering procedure described in subsection \ref{sec:con_model} yields $k_1 = 4$ coefficient pairs $(\alpha, \beta)$ and $k_2 = 4$ distinct $\vartheta$ values, for a total of 13 parameters, 19 fewer than in d-vMEM-SeC.\footnote{For clustering we employ an agglomerative hierarchical algorithm with average linkage criterion,  choosing the number of clusters that provides the largest vertical difference between nodes. To save space we do not show the numerical results of each step of the procedure, available upon request.} 

As shown in Table \ref{tab:clust}, most assets fall into clusters 2 and 3 for $(\alpha,\beta)$ (11 and 15 assets respectively). Cluster 1 includes only AAPL and AMZN, while cluster 4 contains only HON. The clustering of $\vartheta$ coefficients does not closely align with that of $(\alpha,\beta)$. For instance, HON uniquely belongs to cluster 4 in both cases, while AAPL and AMZN - together in cluster 1 for $(\alpha,\beta)$ - are also in cluster 1 for $\vartheta$, but with five additional assets. The overall similarity between the two clusterings is limited, as confirmed by an Adjusted Rand Index (ARI) of 0.38.\footnote{The ARI \citep{ha85} measures the agreement between two clustering schemes. It typically ranges from 0 (random agreement) to 1 (perfect agreement), with rare negative values when the agreement is worse than random.}

\begin{table}[h!]
		\caption{Estimation results for vMEM and vMEM-SeC models (robust standard errors in parentheses).  \label{tab:est}}
		\begin{tabular}{lcccccc} \hline
				Parameter&s-vMEM&d-vMEM&c-vMEM&s-vMEM-SeC&d-vMEM-SeC&c-vMEM-SeC\\
				$\alpha_1$&0.099&0.123&0.118&0.077&0.147&0.132\\
				&(0.004)&(0.010)&(0.007)&(0.004)&(0.018)&(0.012)\\
				$\alpha_2$&&0.097&0.096&&0.068&0.066\\
				&&(0.008)&(0.004)&&(0.016)&(0.004)\\
				$\alpha_3$&&0.102&0.100&      &0.089&0.084\\
				&&(0.008)&(0.004)&&(0.018)&(0.005)\\
				$\alpha_4$& & &&      &0.033&0.037\\
				&& &&&(0.021)&(0.015)\\
				$\beta_1$ &0.877&0.841&0.852&0.903&0.797&0.826\\
				&(0.006)&(0.015)&(0.010)&(0.006)&(0.028)&(0.019)\\
				$\beta_2$&&0.876&0.881&&0.909&0.920\\
				&&(0.011)&(0.006)&&(0.023)&(0.006)\\
				$\beta_3$ &&0.870&0.876      &      &0.879&0.891\\
				&&(0.012)&(0.006)&&(0.027)&(0.007)\\
				$\beta_4$ &&&      &      &0.959&0.953\\
				&&&&&(0.029)&(0.021)\\
				$\delta$&&&&0.074&0.075&0.075\\
				&&&&(0.003)&(0.004)&(0.004)\\
				$\phi$&&&&0.367&0.387&0.391\\	
				&&&&(0.034)&(0.034)&(0.033)\\
				$\vartheta_1$&&&&&0.867&0.884\\	
				&&&&&(0.060)&(0.014)\\
				$\vartheta_2$&&&&&1.083&1.076\\	
				&&&&&(0.049)&(0.010)\\
				$\vartheta_3$&&&&&0.974&0.972\\	
				&&&&&(0.057)&(0.009)\\\hline
		\end{tabular}
\footnotesize{Robust standard errors are computed considering the sandwich estimator of the covariance matrix \citep{White82}. The values corresponding to $\alpha$ and $\beta$ parameters in d-vMEM and values  corresponding to $\alpha$ and $\beta$ parameters in the d-vMEM-SeC,  are the averages and the standard deviations of the 29 corresponding estimates of the parameters belonging to the same clusters obtained for c-vMEM and c-vMEM-SeC respectively (see Table \ref{tab:clust}).}
\end{table}

For c-vMEM, a similar clustering strategy is applied, using only steps 4. and 5. of the algorithm, with $\bm{\nu}_t^*={\bm x}_t$.  In this case, three groups for $(\alpha,\beta)$ are identified. These are shown in Table \ref{tab:clust} (in parentheses). The two sets of clusters for $(\alpha,\beta)$ differ substantially, with an ARI of 0.35. A chi-square test of independence does not reject the null hypothesis at the 1\% level, suggesting that including spillover and co-movement leads to notable changes in the classification of parameters.

Table \ref{tab:est} presents parameter estimates for all six models. For d-vMEM and d-vMEM-SeC, which estimate 58 and 89 parameters respectively, we report average estimates and standard deviations for each cluster identified in Table \ref{tab:clust}. Comparing these to the corresponding estimates in c-vMEM and c-vMEM-SeC provides a check on the clustering quality.

Overall, the differences between d-vMEM and c-vMEM, as well as between d-vMEM-SeC and c-vMEM-SeC, are small. On average, diagonal models exhibit slightly larger $\alpha_i$ values and slightly smaller $\beta_i$ values (with the exception of cluster 4 in SeC models). The largest differences ($\beta_1$ in both cases) are modest: 0.011 for the vMEM and 0.029 for the vMEM-SeC models. Differences among $\alpha_i$ and $\beta_i$ in c-vMEM are minimal, while variation is more pronounced in c-vMEM-SeC (e.g., a 0.095 difference for $\alpha$, 0.127 for $\beta$). The $\vartheta_i$ coefficients in SeC models show well-separated clusters, with only minor differences between the diagonal and clustered models (maximum 0.017 for $\vartheta_1$). Note that $\vartheta_4$, determined by the constraint $\bm{\vartheta}' \bm{\iota}_{29} = 29$, is equal to 1.152. Finally, the co-movement parameters $\delta$ and $\phi$ are consistent across all three SeC specifications.

\begin{table}[h!]
	\centering
	\caption{Likelihood-based criteria and loss functions for competing vMEM and vMEM-SeC models. The bold characters indicate the best index; the stars indicate that corresponding models fall in the best MCS at 5\% significance level.  \label{tab:loss}}
	{\begin{tabular}{l|rrrrrrr} \hline
			Model&Log-Lik.&AIC&BIC&MSE i.s.&Qlike i.s.:&MSE o.o.s.&Qlike o.o.s.\\
			s-vMEM&416608.3&-205.68&-205.68&3.5643 & -3.1952 & 2.4012 &-3.3765$^*$\\
			d-vMEM&416698.5&-205.70&-205.61&3.5569 & -3.1952 & 2.4032& -3.3764$^*$\\
			c-vMEM&416619.2&-205.68&-205.67&3.5608 & -3.1953 & 2.4007 & -3.3766$^*$\\
			s-vMEM-SeC&417245.0&-205.99&-205.99&3.0493$^*$ & -3.2333& \bf{2.3757}$^*$& \bf{-3.3790}$^*$\\
			d-vMEM-SeC&{\bf417441.2}&\bf{-206.05}&-205.91&\bf{3.0211}$^*$&\bf{-3.2348}$^*$ & 2.3872& -3.3779$^*$\\
			c-vMEM-SeC&417336.4&-206.03&{\bf-206.01}& 3.0394$^*$&-3.2341&2.3808$^*$&-3.3782$^*$\\ \hline
	\end{tabular}}
\end{table}

Table \ref{tab:loss} reports model performance using likelihood-based criteria (AIC, BIC) and loss functions (MSE and QLIKE), both in-sample and out-of-sample. Within each model class, AIC slightly favors diagonal models, while BIC (penalizing complexity) prefers more parsimonious ones. All vMEM-SeC models outperform their vMEM counterparts; notably, c-vMEM-SeC achieves the best BIC, highlighting the effectiveness of the clustering algorithm.

Regarding loss functions, vMEM-SeC models consistently achieve lower in-sample MSE and QLIKE values; d-vMEM-SeC performs best in this regard. To formally assess performance differences, we apply the Model Confidence Set (MCS) procedure of \citet{hln2011}, using the Semi-Quadratic statistic with 1,000 bootstrap resamples. At the 5\% level, all vMEM-SeC models are included in the best MCS set for MSE, while d-vMEM-SeC is significantly the best for QLIKE. These results confirm the superior in-sample performance of the SeC models.

Out-of-sample results (last two columns of Table \ref{tab:loss}) also favor SeC models. The s-vMEM-SeC model achieves the lowest MSE and QLIKE values. The strong out-of-sample performance of this more parsimonious model aligns with findings in the literature that simpler models often generalize better in out-of-sample terms \citep[see, e.g.,][]{HT2015, XieYu2020}.  A less obvious result is that once again the SeC models win over all considered alternatives, showing lower loss functions. While only s-vMEM-SeC and c-vMEM-SeC enter the best MCS  in terms of MSE, all models are included for QLIKE 

\begin{table}[h!]
		\caption{Coefficients of each c-vMEM-SeC equation, as in equation (\ref{unimu}); spillover coefficients refer only to the AAPL equation. \label{tab:unicoeff}}
		\begin{tabular}{lcccc} \hline
			asset&($x_{i,t-1}-\bar{x}_i$)&($\ln \mu_{i,t-1}-\bar{x}_i$)&($x_{j,t-1}-\bar{x}_j$)&($\xi_{t-1}$)\\ \hline
			AAPL & 0.143 & 0.826 & - & -0.501 \\ 
			AMGN & 0.093 & 0.891 & 0.008 & -0.517 \\ 
			AMZN & 0.144 & 0.826 & 0.012 & -0.501 \\ 
			AXP & 0.085 & 0.920 & 0.016 & -0.639 \\ 
			BA & 0.080 & 0.920 & 0.013 & -0.577 \\ 
			CAT & 0.096 & 0.891 & 0.012 & -0.517 \\ 
			CRM & 0.096 & 0.891 & 0.011 & -0.568 \\ 
			CSCO & 0.078 & 0.920 & 0.012 & -0.577 \\ 
			CVX & 0.099 & 0.891 & 0.012 & -0.629 \\ 
			DIS & 0.097 & 0.891 & 0.013 & -0.517 \\ 
			GS & 0.098 & 0.891 & 0.013 & -0.568 \\ 
			HD & 0.099 & 0.891 & 0.013 & -0.568 \\ 
			HON & 0.057 & 0.953 & 0.014 & -0.752 \\ 
			IBM & 0.097 & 0.891 & 0.012 & -0.568 \\ 
			INTC & 0.078 & 0.920 & 0.011 & -0.577 \\ 
			JNJ & 0.095 & 0.891 & 0.010 & -0.568 \\ 
			JPM & 0.084 & 0.920 & 0.015 & -0.639 \\ 
			KO & 0.080 & 0.920 & 0.012 & -0.639 \\ 
			MCD & 0.080 & 0.920 & 0.012 & -0.639 \\ 
			MMM & 0.079 & 0.920 & 0.013 & -0.577 \\ 
			MRK & 0.078 & 0.920 & 0.011 & -0.577 \\ 
			MSFT & 0.080 & 0.920 & 0.012 & -0.639 \\ 
			NKE & 0.079 & 0.920 & 0.011 & -0.639 \\ 
			PG & 0.079 & 0.920 & 0.011 & -0.639 \\ 
			TRV & 0.083 & 0.920 & 0.014 & -0.639 \\ 
			UNH & 0.078 & 0.920 & 0.012 & -0.525 \\ 
			V & 0.101 & 0.891 & 0.014 & -0.629 \\ 
			VZ & 0.078 & 0.920 & 0.010 & -0.639 \\ 
			WMT & 0.094 & 0.891 & 0.010 & -0.517 \\ 
		\end{tabular}
\end{table}

\subsection{Contribution of the factors}\label{sec:fac}
The vMEM-SeC clearly outperforms the standard vMEM on the dataset under study, even when considering the simplest scalar version. Moreover, for each HLR included in the dataset, it is possible to capture both spillover effects and co-movements, characterized by different coefficients, as shown in equation (\ref{unimu}).  Focusing on the c-vMEM-SeC, Table \ref{tab:unicoeff} reports the coefficients of the regressors in the right-hand side of equation (\ref{unimu}), presented equation by equation. Regarding the spillover effect coefficients (third column of Table \ref{tab:unicoeff}), we only display those for the first ticker, which refers to AAPL. The table highlights the model's ability to allow for different coefficients in each equation within a parsimonious structure involving just 13 parameters, effectively separating the GARCH-type dynamics of each asset from the spillover and co-movement components. 

\begin{figure}[h!]
	\caption{Daily log HLR (black lines) of 29 components of the Dow Jones Industrial (DJI) and estimated volatility (blue lines) with c-vMEM-SeC belonging to group 3 in the ($\alpha,\beta$) classification (see Table \ref{tab:clust}). In sample period:  19 March 2008 -- 22 April 2024}
	\begin{subfigure}[h!]{0.32\textwidth}
		\centering		
		\includegraphics[page=1, height=3.3cm, width=4.5cm]{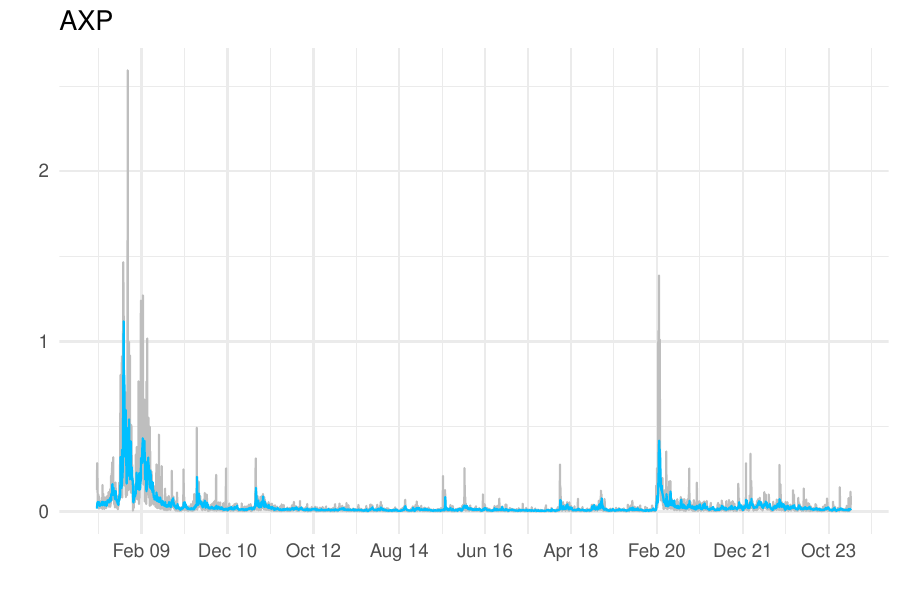}  
	\end{subfigure}
	\hfill 
	\begin{subfigure}[h!]{0.32\textwidth}
		\centering
		\includegraphics[page=2, height=3.3cm, width=4.5cm]{cluster_HL3.pdf}
	\end{subfigure}   
	\hfill
	\begin{subfigure}[h!]{0.32\textwidth}
		\centering		
		\includegraphics[page=3,height=3.3cm, width=4.5cm]{cluster_HL3.pdf} 
	\end{subfigure} 
	\vfill	 \begin{subfigure}[h!]{0.32\textwidth}
		\centering
		\includegraphics[page=4, height=3.3cm, width=4.5cm]{cluster_HL3.pdf}
	\end{subfigure} 
	\hfill 
	\begin{subfigure}[h!]{0.32\textwidth}
		\centering
		\includegraphics[page=5, height=3.3cm, width=4.5cm]{cluster_HL3.pdf}
	\end{subfigure} 
	\hfill\begin{subfigure}[h!]{0.32\textwidth}
		\centering
		\includegraphics[page=6, height=3.3cm, width=4.5cm]{cluster_HL3.pdf}
	\end{subfigure} 
	\vfill
	\begin{subfigure}[h!]{0.32\textwidth}
		\centering		
		\includegraphics[page=7, height=3.3cm, width=4.5cm]{cluster_HL3.pdf}  
	\end{subfigure}
	\hfill 
	\begin{subfigure}[h!]{0.32\textwidth}
		\centering
		\includegraphics[page=8, height=3.3cm, width=4.5cm]{cluster_HL3.pdf}
	\end{subfigure}   
	\hfill
	\begin{subfigure}[h!]{0.32\textwidth}
		\centering		
		\includegraphics[page=9,height=3.3cm, width=4.5cm]{cluster_HL3.pdf} 
	\end{subfigure} 
	\vfill	 \begin{subfigure}[h!]{0.32\textwidth}
		\centering
		\includegraphics[page=10, height=3.3cm, width=4.5cm]{cluster_HL3.pdf}
	\end{subfigure} 
	\hfill 
	\begin{subfigure}[h!]{0.32\textwidth}
		\centering
		\includegraphics[page=11, height=3.3cm, width=4.5cm]{cluster_HL3.pdf}
	\end{subfigure} 
	\hfill\begin{subfigure}[h!]{0.32\textwidth}
		\centering
		\includegraphics[page=12, height=3.3cm, width=4.5cm]{cluster_HL3.pdf}
	\end{subfigure} 
	\vfill
	\begin{subfigure}[h!]{0.32\textwidth}
		\centering		
		\includegraphics[page=13, height=3.3cm, width=4.5cm]{cluster_HL3.pdf}  
	\end{subfigure}
	\hfill 
	\begin{subfigure}[h!]{0.32\textwidth}
		\centering
		\includegraphics[page=14, height=3.3cm, width=4.5cm]{cluster_HL3.pdf}
	\end{subfigure}   
	\hfill
	\begin{subfigure}[h!]{0.32\textwidth}
		\centering		
		\includegraphics[page=15,height=3.3cm, width=4.5cm]{cluster_HL3.pdf} 
	\end{subfigure} 
	\label{fig:estimHL1}       
\end{figure}

\begin{figure}[h!]
	\caption{Daily log HLR (black lines) of 29 components of the Dow Jones Industrial (DJI) and estimated volatility (blue lines) with c-vMEM-SeC belonging to groups 1, 2, 4 in the ($\alpha,\beta$) classification (see Table \ref{tab:clust}). In sample period:  19 March 2008 -- 22 April 2024}
	\begin{subfigure}[h!]{0.32\textwidth}
		\centering		
		\includegraphics[page=1, height=3.3cm, width=4.5cm]{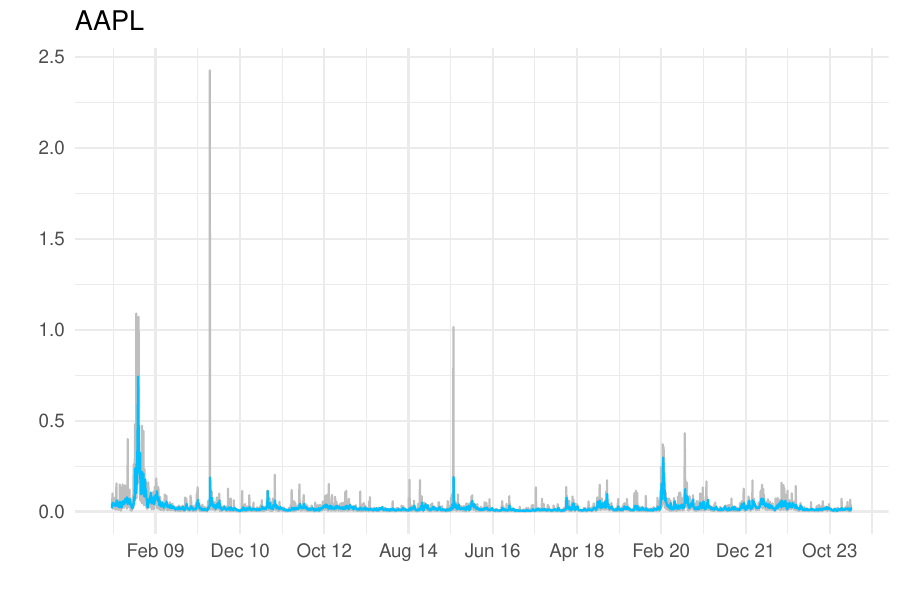}  
	\end{subfigure}
	\hfill 
	\begin{subfigure}[h!]{0.32\textwidth}
		\centering
		\includegraphics[page=2, height=3.3cm, width=4.5cm]{cluster_HL124.pdf}
	\end{subfigure}   
	\hfill
	\begin{subfigure}[h!]{0.32\textwidth}
		\centering		
		\includegraphics[page=3,height=3.3cm, width=4.5cm]{cluster_HL124.pdf} 
	\end{subfigure} 
	\vfill	 \begin{subfigure}[h!]{0.32\textwidth}
		\centering
		\includegraphics[page=4, height=3.3cm, width=4.5cm]{cluster_HL124.pdf}
	\end{subfigure} 
	\hfill 
	\begin{subfigure}[h!]{0.32\textwidth}
		\centering
		\includegraphics[page=5, height=3.3cm, width=4.5cm]{cluster_HL124.pdf}
	\end{subfigure} 
	\hfill\begin{subfigure}[h!]{0.32\textwidth}
		\centering
		\includegraphics[page=6, height=3.3cm, width=4.5cm]{cluster_HL124.pdf}
	\end{subfigure} 
	\vfill
	\begin{subfigure}[h!]{0.32\textwidth}
		\centering		
		\includegraphics[page=7, height=3.3cm, width=4.5cm]{cluster_HL124.pdf}  
	\end{subfigure}
	\hfill 
	\begin{subfigure}[h!]{0.32\textwidth}
		\centering
		\includegraphics[page=8, height=3.3cm, width=4.5cm,]{cluster_HL124.pdf}
	\end{subfigure}   
	\hfill
	\begin{subfigure}[h!]{0.32\textwidth}
		\centering		
		\includegraphics[page=9,height=3.3cm, width=4.5cm]{cluster_HL124.pdf} 
	\end{subfigure} 
	\vfill	 \begin{subfigure}[h!]{0.32\textwidth}
		\centering
		\includegraphics[page=10, height=3.3cm, width=4.5cm]{cluster_HL124.pdf}
	\end{subfigure} 
	\hfill 
	\begin{subfigure}[h!]{0.32\textwidth}
		\centering
		\includegraphics[page=11, height=3.3cm, width=4.5cm]{cluster_HL124.pdf}
	\end{subfigure} 
	\hfill\begin{subfigure}[h!]{0.32\textwidth}
		\centering
		\includegraphics[page=12, height=3.3cm, width=4.5cm]{cluster_HL124.pdf}
	\end{subfigure} 
	\vfill
	\begin{subfigure}[h!]{0.32\textwidth}
		\centering		
		\includegraphics[page=13, height=3.3cm, width=4.5cm]{cluster_HL124.pdf}  
	\end{subfigure}
	\hfill 
	\begin{subfigure}[h!]{0.32\textwidth}
		\centering
		\includegraphics[page=14, height=3.3cm, width=4.5cm]{cluster_HL124.pdf}
	\end{subfigure}   
	\label{fig:estimHL2}
\end{figure}

\begin{figure}[h!]
	\caption{Common component ($exp(\xi_t)$) of the 29 HLR series belonging to the DJI. In sample period:  19 March 2008 -- 22 April 2024}\label{fig:common}
	\centering		
	\includegraphics[scale=0.53]{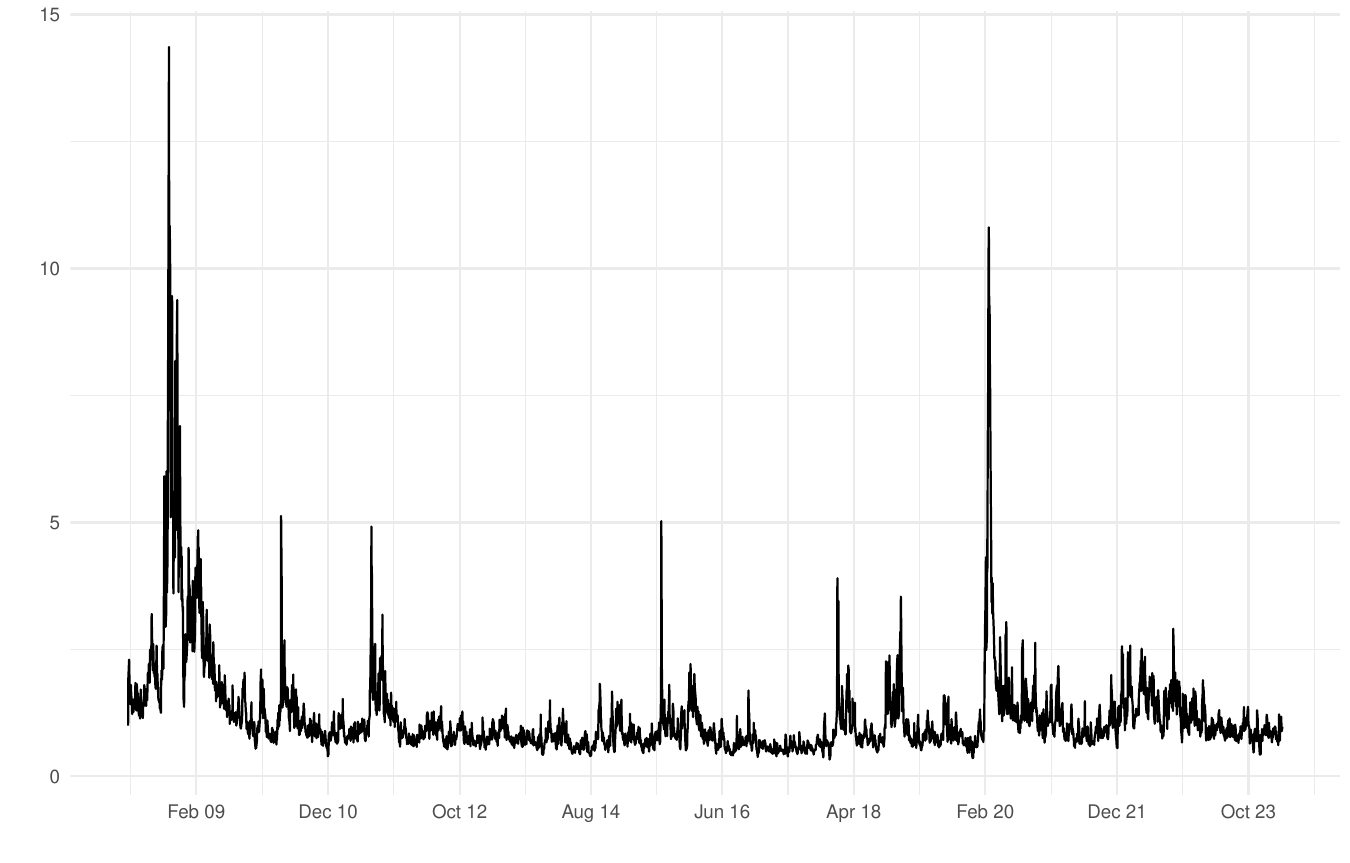}  
\end{figure}

\begin{figure}[h!]
	\caption{Estimated volatility ($\mu_{i,t}$, gray continuous line) and idiosyncratic component ($exp(\varsigma_{i,t})$,  blue continuous  line) of AAPL. In sample period:  19 March 2008 -- 22 April 2024}\label{fig:mu_sigma_AAPL}
	\centering		
	\includegraphics[scale=0.53]{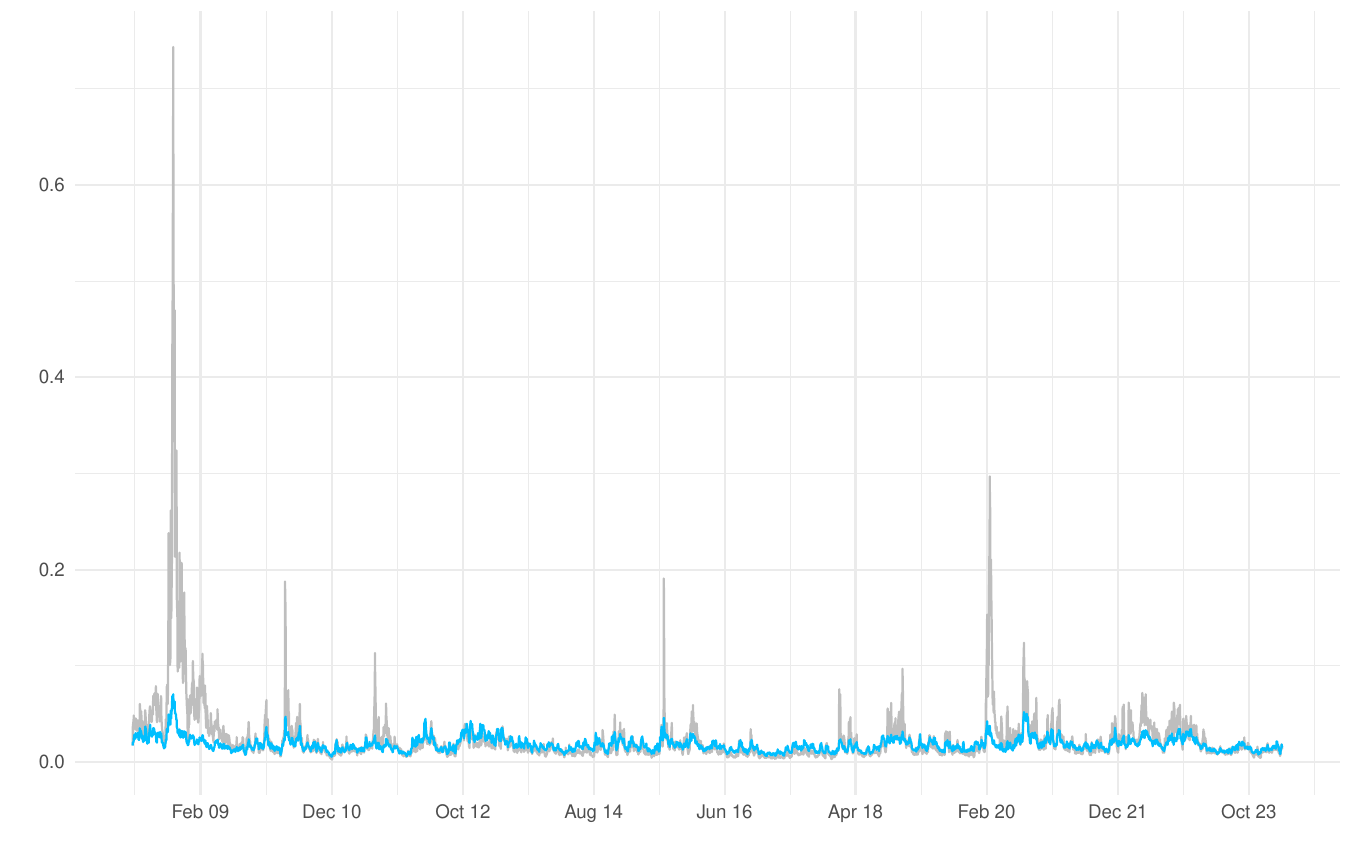}  
\end{figure}

More insightful is the graphical analysis of the components.
Figures \ref{fig:estimHL1} and \ref{fig:estimHL2} display the 29 estimated volatilities $\mu_{i,t}$ (blue lines), obtained from the c-vMEM-SeC, overlaid on the corresponding 29 HLR series (black lines).\footnote{To facilitate visual inspection, we present in Figure \ref{fig:estimHL1} the series belonging to cluster 3 (as classified by the $\alpha,\beta$ parameters), and the remaining series in Figure \ref{fig:estimHL2}.} Notably, despite the small number of estimated parameters, the dynamics of each series is highly flexible, effectively adapting to the diverse behavior of the HLR series. This flexibility is made possible by the idiosyncratic components in the vector $\bm{\varsigma}_t$, while shared dynamics are captured by $\xi_t$, which enters each series through different coefficients in $\bm{\theta}$. 

The common component is shown in Figure \ref{fig:common}; its dynamics clearly reflect jumps corresponding to shocks that affected the economic behavior of all financial assets, such as the collapse of Lehman Brothers in September 2008, the flash crash in May 2010, the weakening of the US economy and the intensification of the European debt crisis in August 2011, the stock market sell-off of 2015-16, the financial crisis of 2018-19, and the COVID-19 crash in 2020.

The contribution of the co-movement to volatility can be appreciated by comparing the estimated volatility with its idiosyncratic component. Figure \ref{fig:mu_sigma_AAPL} shows this comparison for the ticker AAPL. The idiosyncratic component generally tracks the volatility dynamics with a profile with low variability; during crises, its magnitude is amplified by the common factor, which is multiplied by it through the coefficient $\theta_1$.

\newpage

\section{Concluding Remarks}\label{sec:concl}
A significant strand of the statistical and econometric literature focuses on incorporating the influence of other markets' dynamics into volatility models, leading to the development of various multivariate approaches, particularly those addressing spillover and contagion effects \citep[see, for example,][]{Pericoli_Sbracia:2003}. Most of these approaches construct ad hoc models to capture such interdependencies, often within a VAR framework \citep{Gallo_Otranto:2008, Diebold_Yilmaz:2012}. The vector Multiplicative Error Model (vMEM) emerges as a promising candidate for modeling spillover effects, as it integrates flexible GARCH-type dynamics for volatility modeling and, through the composite MEM extension \citep{mem_hbv_2012}, allows for the specification of multiple latent components. We exploit this latter feature to decompose the expected conditional variance into two parts: a GARCH-type component, which accounts for idiosyncratic volatility dynamics, and an unobservable common component, which  reflects typical synchronization patterns observed in financial markets and volatility spillover effects \citep{ForbesRigobon2002}. 

To enable the modeling of a high-dimensional system of volatility series, we propose a model-based clustering approach that substantially reduces the number of free parameters. In the presented empirical application, only 13 parameters are estimated, compared to the 523 parameters required in the fully parameterized model (including covariance terms), representing a reduction of approximately 97.5\%.

The empirical illustration, based on a limited set of assets (29 components of the Dow Jones Industrial Average), enables a direct comparison between different parameterizations of the proposed model and the baseline vMEMs. In particular, we compare the clustered parameterization (with 13 estimated parameters) to the diagonal benchmark parameterization, which includes 89 parameters and remains estimable, albeit at a higher computational cost. The results indicate comparable performance between the two specifications, especially in terms of forecasting accuracy. This suggests that the clustered specification is a viable alternative for high-dimensional applications where the diagonal specification becomes computationally prohibitive.

The proposed model is well-suited for forecasting purposes and for identifying the directionality of spillover effects, thereby enabling the detection of market clusters with varying degrees of shock transmission-so-called dominant markets \citep[see][]{Gallo_Otranto:2008, Otranto:Gargano:2014}. Future research could explore the integration of alternative methods for identifying co-movements, such as Independent Component Analysis \citep[see, e.g.,][]{Hyvarinen:Karhunen:Oja:2001}. Moreover, it would be of interest to investigate the inclusion of alternative parameter grouping strategies, such as the fuzzy clustering approach proposed by \citet{Cerqueti_et_al:2023}, and to examine the potential of the model for classification tasks.

The \textsf{R} codes \citep{R:2018} used to implement the analyses in Section \ref{sec:apps} were developed by the authors.

\backmatter

\bmhead{Acknowledgements}

Edoardo Otranto acknowledges financial support from the Italian PRIN 2022 grant ``Methodological and computational issues in large-scale time series models for economics and finance'' (20223725WE). Luca Scaffidi Domianello acknowledges financial support from PNRR MUR project PE0000013-FAIR.

\section*{Statements and Declarations}

\noindent
The authors report there are no competing interests to declare.

\newpage

\begin{appendices}

\section{Expectation Targeting Estimation}\label{targeting}
Expectation targeting is a two-step estimation procedure in which the intercept term is expressed as a function of the unconditional mean. The latter is estimated using sample moments in the first step and subsequently substituted into the log-likelihood function during the second step, where the remaining model parameters are estimated via maximum likelihood \citep[see, for example,][]{mem_hbv_2012}. This approach is analogous to the variance targeting technique employed in the estimation of GARCH models \citep[see, e.g.,][]{Engle:Mezrich:1996}.

Consider a log-vMEM process for the series $\left\{\bm{y}_t \right\}$, with $\left\{\bm{x}_t\right\}$ denoting the logarithm of $\bm{y}_t$:
\begin{align}
	\begin{split}
		{\bm y}_t&={\bm \mu}_t \odot {\bm \varepsilon}_t   \qquad {\bm \varepsilon}_t \sim {\ln N}({\bm m},{\bm V})  \\
		\ln {\bm \mu}_t&= \bm{\omega}+ {\bm A} {\bm x}_{t-1}+{\bm B}  {\ln\bm{\mu} }_{t-1} 
	\end{split}\label{svmem_sc}
\end{align} 
where the condition $\bm m=-diag(\bm{V})/2$  ensures that the expectation of $\bm{\varepsilon}_t$ is a vector of ones.  Under the assumption of stationarity for $\bm{x}_t$, the first equation in~\eqref{svmem_sc} implies $E(\ln\bm{\mu}_t)=E(\bm{x}_t)+diag(\bm{V})/2$.\footnote{Note that $E(\ln\bm{\varepsilon}_t)$ depends on the distributional assumption for the innovations. Under the log-normal assumption with $\bm{m}=-diag(\bm{V})/2$ we obtain $E(\ln\bm{\varepsilon}_t)=-diag(\bm{V})/2$, yielding unit mean for $\bm{\varepsilon}_t$.} The unconditional mean $E(\bm{x}_t)$ can be estimated by the sample average $\bar{\bm{x}}=T^{-1}\sum_{t=1}^T\bm{x}_t$. Taking expectations on both sides of the second equation in~\eqref{svmem_sc} yields: 
\begin{align*}
	E(\ln {\bm \mu}_t)&= \bm{\omega}+ {\bm A} E({\bm x}_{t-1})+{\bm B}  E({\ln\bm{\mu} }_{t-1}) \\
	\bar{\bm{x}}+diag(\bm{V})/2&= \bm{\omega} + {\bm A}\bm{x}+{\bm B}  (\bar{\bm{x}}+diag(\bm{V})/2) \\
	\bm{\omega}&=(\bm{I}_n-{\bm A}-{\bm B})\bar{\bm{x}} + (\bm{I}_n-{\bm B})diag(\bm{V})/2
\end{align*}\label{Esvmem_sc}
Substituting this expression into~\eqref{svmem_sc} yields the expectation-targeted specification:
\begin{equation}
	\ln {\bm \mu}_t= (\bm{I}_n-{\bm A}-{\bm B})\bar{\bm{x}} + (\bm{I}_n-{\bm B})diag(\bm{V})/2+ {\bm A} {\bm x}_{t-1}+{\bm B}  {\ln\bm{\mu} }_{t-1} 
\end{equation} 
This reparameterization removes $\bm{\omega}$ from the optimization procedure, improving numerical stability and reducing the number of parameters to be estimated. Since the matrix $\bm{V}$ is updated at each iteration using the sample covariance matrix of the log-residuals $\ln \hat{\bm{\varepsilon}}_t$ obtained from the previous step, numerical consistency is preserved throughout the estimation process.

A similar reasoning applies to the specification in~\eqref{vmem_sc}, noting that $E(\bm{x}_t) = E(\bm{\nu}_t)$ due to the mean-zero assumption for the innovation term $\xi_t$.

\section{The distance between ARMA(1,1) processes}\label{app:dist}
Consider two ARMA(1,1) processes defined as:
\begin{equation}
	x_{i,t}=\varphi_i x_{i,t-1}+\eta_{i,t}-\psi_i \eta_{i,t-1} \qquad i=1,2 \label{arma}
\end{equation}
where $\eta_{i,t}$ denotes a white noise process.

Using the lag operator $L$ (such that $L^j x_{i,t} = x_{i,t-j}$), equation~\eqref{arma} can be rewritten using lag polynomials as:
\begin{equation}
	(1-\varphi_i L) x_{i,t}=(1-\psi_i L)\eta_{i,t} \label{armalag}
\end{equation}

The AR distance proposed by \cite{Piccolo_1990} is defined as the Euclidean distance between the infinite-order autoregressive (AR) coefficients of two stochastic processes. Denoting by $\pi_{i,j}$ the $j$-th AR coefficient of process $i$ ($i = 1,2$), the AR distance is:
\begin{equation}
	d_{AR}=\left[\sum_{j=1}^{\infty} (\pi_{1,j}-\pi_{2,j})^2 \right]^{1/2} \label{d_ar}
\end{equation}
This metric has been extensively employed to assess whether two models exhibit similar dynamic behavior \citep[see, e.g.,][]{Otranto:2010, Otranto:Gargano:2014}. For ARMA(1,1) processes, this distance can be derived explicitly.

Assuming invertibility, equation~\eqref{armalag} can be written as:
\begin{equation}
	(1-\varphi_i L)(1-\psi_i L)^{-1} x_{i,t}=\eta_{i,t} \label{arlag}
\end{equation}
Under the condition $|\psi_i| < 1$, the inverse of the MA operator can be expressed as a power series, yielding:
\begin{equation}
	(1-\varphi_i L)\sum_{j=1}^{\infty}(\psi_i L)^j x_{i,t}=\eta_{i,t} \label{arinf}
\end{equation}
From this expression, the $j$-th AR coefficient is given by:
\begin{equation}
	\pi_{i,j}=\varphi_i\psi_i^{j-1}-\psi_i^j \label{pi}
\end{equation}

Substituting equation~\eqref{pi} into~\eqref{d_ar}, the AR distance becomes:
\begin{align}
	\begin{array}{l}
		\left[\sum_{j=1}^{\infty} (\varphi_1 \psi_1^{j-1}-\psi_1^j)^2+(\varphi_2 \psi_2^{j-1}-\psi_2^j)^2-2(\varphi_1 \psi_1^{j-1}-\psi_1^j)(\varphi_2 \psi_2^{j-1}-\psi_2^j) \right]^{1/2}=\\
		\left[\sum_{j=1}^{\infty}\left(\varphi_1^2 \psi_1^{2(j-1)}+\psi_1^{2j}-2 \varphi_1 \psi_1^{2j-1}+\varphi_2^2 \psi_2^{2(j-1)}+\psi_2^{2j}-2 \varphi_2 \psi_2^{2j-1}-2 \varphi_1\varphi_2(\psi_1\psi_2)^{j-1}+\right. \right.\\
		\left. \left. 2\varphi_2\psi_1(\psi_1\psi_2)^{j-1}+2\varphi_1\psi_2(\psi_1\psi_2)^{j-1}-2(\psi_1\psi_2)^j \right) \right]^{1/2}
	\end{array} \label{dist_exp}
\end{align}

Under invertibility, $\psi_1^2<1$, $\psi_2^2<1$, $\psi_1 \psi_2<1$; labeling with $\kappa$ one of the three previous functions of $\psi_1$ and/or $\psi_2$, each $\sum_{j=1}^{\infty} \kappa^{j-1}=\frac{1}{1-\kappa}$ and $\sum_{j=1}^{\infty} \kappa^{j}=\frac{\kappa}{1-\kappa}$. Using these identities, equation~\eqref{dist_exp} simplifies to:

\begin{align}
	\begin{array}{l}
		\left[\frac{\varphi_1^2}{1-\psi_1^2}+\frac{\psi_1^2}{1-\psi_1^2}-2\frac{\varphi_1 \psi_1}{1- \psi_1^2}+\frac{\varphi_2^2}{1-\psi_2^2}+\frac{\psi_2^2}{1-\psi_2^2}-2\frac{\varphi_2 \psi_2}{1- \psi_2^2}-2\frac{\varphi_1 \varphi_2}{1- \psi_1 \psi_2}	2\frac{\varphi_2 \psi_1}{1- \psi_1 \psi_2}+ \right.\\
		\left. 	2\frac{\varphi_1 \psi_2}{1- \psi_1 \psi_2}-2\frac{\psi_1 \psi_2}{1- \psi_1 \psi_2}\right]^{1/2}
	\end{array}
\end{align}

This can be further compacted into the following closed-form expression:
\begin{equation}
	d_{ARMA}=\left[\frac{(\varphi_1-\psi_1)^2}{1-\psi_1^2}+\frac{(\varphi_2-\psi_2)^2}{1-\psi_2^2}-2\frac{(\varphi_1-\psi_1)(\varphi_2-\psi_2)}{1-\psi_1\psi_2} \right]^{1/2}
	\label{d_arma}
\end{equation}

Starting from the final equation in~\eqref{unimem}, and adding and subtracting $\nu_{i,t}$ and $\beta_i \nu_{i,t-1}$ on the right-hand side, we can algebraically derive an ARMA(1,1) representation for $\nu_{i,t}$:
\begin{equation}
	\nu_{i,t}=c_i+ (\alpha_i+\beta_i) \nu_{i,t-1}-\beta_i(\nu_{i,t-1}- \varsigma_{i,t-1})+ (\nu_{i,t}-\varsigma_{i,t})	
\end{equation}
where $c_i= (1 - \alpha_i- \beta_i) \bar{x_i} + (1 - \beta_i)v_{ii}/2$. Hence, the ARMA distance given in~\eqref{d_arma} can be applied and leads to the form presented in~\eqref{armadist}.

A similar approach for defining distances was proposed in \citet{Otranto:2010} for DCC models.



\end{appendices}


\bibliography{biblio}

@Manual{R:2018,
	title = {\textsf{R}: A Language and Environment for Statistical Computing},
	author = {{\textsf{R} Core Team}},
	organization = {\textsf{R} Foundation for Statistical Computing},
	address = {Vienna, Austria},
	year = {2018}
	}

@book{Hyvarinen:Karhunen:Oja:2001,
  title={Independent component analysis},
  author={Hyvärinen, Aapo and Karhunen, Juha and Oja, Erkki},
  year={2001},
  publisher={{J}ohn {W}iley \& {S}ons},
address={{N}ew {Y}ork}
}

@article{HT2015,
	author = {Hansen, P. R. and Timmerman, A.},
	journal = {Journal of Business and Economic Statistics},
	pages = {17-21},
	title = {Comment
on {F.X.} {D}iebold: comparing predictive
accuracy: twenty years later. A personal
perspective on the use and abuse of
Diebold-Mariano Tests.},
	volume = { 33},
	year = {2015}}

@article{XieYu2020,
	author = {Xie, H. and Yu, C. .},
	journal = {Finance Research Letters},
	pages = {101221},
	title = {Realized {GARCH} models: Simpler is better},
	volume = { 33},
	year = {2020}}

@article{Laurent:Rombouts:Violante:2012,
  title={On the forecasting accuracy of multivariate {GARCH} models},
  author={Laurent, Sébastien and Rombouts, Jeroen V.K. and Violante, Francesco},
  journal={Journal of Applied Econometrics},
  volume={27},
  pages={934--955},
  year={2012},
  publisher={Wiley}
}

@article{Cipollini:Gallo:2019,
  title={Modeling euro stoxx 50 volatility with common and market-specific components},
  author={Cipollini, Fabrizio and Gallo, Giampiero M},
  journal={Econometrics and Statistics},
  volume={11},
  pages={22--42},
  year={2019},
  publisher={Elsevier}
}

@article{hln2011,
author = {Hansen, Peter R. and Lunde, Asger and Nason, James M.},
title = {The Model Confidence Set},
journal = {Econometrica},
volume = {79},
number = {2},
pages = {453-497},
year = {2011}
}

@article{Chai2020,
	author = {Chai, Shanglei and Zhang, Zhen and Du, Mo and Jiang, Lei},
title = {Volatility Similarity and Spillover Effects in G20 Stock Market Comovements: An ICA-Based ARMA-APARCH-M Approach},
journal = {Complexity},
volume = {2020},
number = {1},
year = {2020}
}

@book{Mardia:1979,
  title={Multivariate analysis},
  author={Mardia, K.V. and Kent, J. T. and Bibby, J.M.},
  year={1979},
  publisher={Academic Press},
 address={London}
}

@article{Cerqueti_et_al:2023,
	author = {Cerqueti, R. and D'Urso, P. and Mattera, R. and Vitale, V.},
	journal = {Annals of Operations Research},
	title = {Fuzzy clustering of financial time series based on volatility spillovers},
	year = {2023}
	}

@article{Gallo_Otranto:2008,
	title = {Volatility spillovers, interdependence and comovements: A {M}arkov {S}witching approach},
	journal = {Computational Statistics \& Data Analysis},
	volume = {52},
	pages = {3011-3026},
	year = {2008},
	author = {Giampiero M. Gallo and Edoardo Otranto},
}

@article{Pericoli_Sbracia:2003,
	title={A Primer on Financial Contagion},
	author={Marcello Pericoli and Massimo Sbracia},
	journal={Journal of Economic Surveys},
	year={2003},
	volume={17},
	pages={571-608},
	}

@article{Diebold_Yilmaz:2012,
	title = {Better to give than to receive: Predictive directional measurement of volatility spillovers},
	journal = {International Journal of Forecasting},
	volume = {28},
	pages = {57-66},
	year = {2012},
	author = {Francis X. Diebold and Kamil Yilmaz},
}

@article{Otranto:Gargano:2014,
	title={Financial clustering in presence of dominant markets},
	author={Edoardo Otranto and Romana Gargano},
	journal={Advances in Data Analysis and Classification},
	year={2014},
	volume={9},
	pages={315-339},
}

@article{Otranto:2010,
	author       = {Edoardo Otranto},
	title        = {Identifying financial time series with similar dynamic conditional
	correlation},
	journal      = {Computational Statistics \& Data Analysis},
	volume       = {54},
	pages        = {1--15},
	year         = {2010},
}

@article{Piccolo_1990,
title = {A DISTANCE MEASURE FOR CLASSIFYING {ARIMA} MODELS},
journal = {Journal of Time Series Analysis},
volume = {11},
pages = {153-164},
year = {1990},
author = {Domenico Piccolo},
}

@Inbook{Chou:Chou:Liu:2015,
	author={Chou, Ray Yeutien
	and Chou, Hengchih
	and Liu, Nathan},
	editor={Lee, Cheng-Few
	and Lee, John C.},
	title={Range Volatility: A Review of Models and Empirical Studies},
	bookTitle={Handbook of Financial Econometrics and Statistics},
	year={2015},
	publisher={Springer New York},
	address={New York, NY},
	pages={2029--2050},
}

@article{Atak_Kapetanios_2013,
title = {A factor approach to realized volatility forecasting in the presence of finite jumps and cross-sectional correlation in pricing errors},
journal = {Economics Letters},
volume = {120},
pages = {224-228},
year = {2013},
author = {Alev Atak and George Kapetanios},
}

@Article{Tailor:Xu:2017,
  author  = {Taylor, N. and Xu, Y.},
  title   = {The logarithmic vector multiplicative error model: an application to high frequency {NYSE} stock data},
  journal = {Quantitative Finance},
  year    = {2017},
  volume  = {17},
  pages   = {1021-1035},
 }

@Article{Parkinson1980,
  author  = {Parkinson, Michael},
  title   = {The Extreme Value Method for Estimating the Variance of the Rate of Return},
  journal = {The Journal of Business},
  year    = {1980},
  volume  = {53},
  pages   = {61-65},
 }

@Article{Cipollini:Engle:Gallo:2013,
  author  = {Cipollini, Fabrizio and Engle, Robert F. and Gallo, Giampiero M.},
  title   = {SEMIPARAMETRIC VECTOR {MEM}},
  journal = {Journal of Applied Econometrics},
  year    = {2013},
  volume  = {28},
  pages   = {1067-1086},
}

@Article{Engle:Mezrich:1996,
	author  = {Engle, Robert F. and Mezrich, J.},
	title   = {{GARCH} for Groups},
	journal = {Risk},
	year    = {1996},
	volume  = {9},
	pages   = {36-40},
}

@article{Otranto:2015,
  title={Capturing the spillover effect with multiplicative error models},
  author={Otranto, Edoardo},
  journal={Communications in Statistics-Theory and Methods},
  volume={44},
  pages={3173--3191},
  year={2015},
  publisher={Taylor \& Francis}
}

@article{Cipollini:Engle:Gallo:2017,
  AUTHOR = {Cipollini, Fabrizio and Engle, Robert F. and Gallo, Giampiero M.},
  TITLE = {Copula–Based {vMEM} Specifications versus Alternatives: The Case of Trading Activity},
  JOURNAL = {Econometrics},
  VOLUME = {5},
  YEAR = {2017},
  NUMBER = {2},
  ARTICLE-NUMBER = {16}
}

@article{BN-etal2008,
	author = {Barndorff-Nielsen, O. and Hansen, P.R. and Lunde, A. and Shephard, N.},
	journal = {Econometrica},
	pages = {1481--1536},
	title = {Designing realised kernels to measure the ex-post variation of equity prices in the presence of noise},
	volume = {76},
	year = {2008}}

@article{ha85,
	author = {Hubert, L. and Arabie, P.},
	journal = {Journal of Classification},
	pages = {193--218},
	title = {Comparing partitions},
	volume = {2},
	year = {1985}}

@article{ForbesRigobon2002,
	author = {Forbes, J. K. and Rigobon, R.},
	journal = {Journal of Finance},
	pages = {2223-2261},
	title = {No Contagion, only Interdependence: Measuring Stock Market Comovement},
	volume = {57},
	year = {2002}}

@article{AndersenBoll98,
	author = {Andersen, T.G. and Bollerslev, T.},
	journal = {International Economic Review},
	pages = {885-905},
	title = {Answering the Skeptics: Yes, Standard Volatility Models do Provide Accurate Forecasts},
	volume = {39},
	year = {1998}}

@article{ABDL01_Econometrica,
	author = {Andersen, T.G. and Bollerslev, T. and Diebold, F.X. and Labys, P.},
	journal = {Econometrica},
	pages = {579-625},
	title = {Modeling and Forecasting Realized Volatility},
	volume = {71},
	year = {2003}}

@article{BLRsurvey06,
	author = {Bauwens, L. and Laurent, S. and Rombouts, J.V.K.},
	journal = {Journal of Applied Econometrics},
	pages = {79-109},
	title = {Multivariate {GARCH} Models: A Survey},
	volume = {21},
	year = {2006}}

@article{Bollerslev86,
	author = {Bollerslev, Tim},
	journal = {Journal of Econometrics},
	pages = {307-327},
	title = {Generalized Autoregressive Conditional Heteroskedasticity},
	volume = {31},
	year = {1986}}

@incollection{mem_hbv_2012,
	author = {Brownlees, Christian~T. and Cipollini, Fabrizio and Gallo, Giampiero~M.},
	booktitle = {Handbook of Volatility Models and Their Applications},
	editor = {Bauwens, Luc and Hafner, Christian and Laurent, S\'ebastien},
	publisher = {Wiley \& Sons, Inc},
 address		= {New Jersey},
        pages = {223-247},
	title = {Multiplicative Error Models},
	year = {2012}}

@article{Corsi_09,
	author = {F. Corsi},
	journal = {Journal of Financial Econometrics},
	pages = {174-196},
	title = {A simple approximate long-memory model of realized volatility},
	volume = {7},
	year = {2009}}

@article{EngleDCC02,
	author = {Engle, Robert F.},
	journal = {Journal of Business \& Economic Statistics},
	pages = {339-350},
	title = {Dynamic Conditional Correlation - a Simple Class of Multivariate {GARCH} Models},
	volume = {20},
	year = {2002}}

@article{Engle_Kroner95,
	author = {Engle, Robert F. and Kroner, F. K.},
	journal = {Econometric Theory},
	pages = {122-150},
	title = {Multivariate Simultaneous Generalized {ARCH}},
	volume = {11},
	year = {1995}}

@article{Engle_NewF_2002,
	author = {Robert F. Engle},
	journal = {Journal of Applied Econometrics},
	pages = {425-446},
	title = {New frontiers for {ARCH} models},
	volume = {17},
	year = {2002}}

@article{EngleGallo_2006,
	author = {Robert F. Engle and Giampiero M. Gallo},
	journal = {Journal of Econometrics},
	pages = {3-27},
	title = {A multiple indicators model for volatility using intra-daily data},
	volume = {131},
	year = {2006}}

@article{White82,
	author = {White, H.},
	journal = {Econometrica},
	pages = {1-25},
	title = {Maximum Likelihood Estimation of Misspecified Models},
	volume = {50},
	year = {1982}}

\end{document}